\documentstyle[prd,aps,epsf,floats,amsfonts,amssymb,amsmath]{revtex}

\newcommand{\bea}{\begin{eqnarray}}\newcommand{\eea}{\end{eqnarray}}

\newcommand{\brr}{\begin{array}}\newcommand{\err}{\end{array}}
\newcommand{\bit}{\begin{itemize}}\newcommand{\eit}{\end{itemize}}
\newcommand{\ben}{\begin{enumerate}}\newcommand{\een}{\end{enumerate}}

\def\lan{\langle}
\def\lf{\left}

\def\non{\nonumber}
\def\ran{\rangle}
\def\rar{\rightarrow}
\def\ri{\right}
\def\wti{\widetilde}
\def\al{\alpha}\def\bt{\beta}
\def\de{\delta}\def\De{\Delta}
\def\te{\theta}

\def\si{\sigma}
\def\om{\omega}

\def\CP{{_{CP}}}
\def\T{{_{T}}}
\begin{document}
%
\title{Comment on  ``Remarks on flavor-neutrino propagators and
oscillation formulae'' }


\author{Massimo Blasone${}^{a,b}$, Antonio Capolupo${}^{b}$  and Giuseppe
Vitiello${}^{b}$
\thanks{e-mail: m.blasone@ic.ac.uk,
capolupo@sa.infn.it,
vitiello@sa.infn.it}
}

\address{${}^{a}$ Blackett Laboratory, Imperial College, Prince Consort
Road, \\ London SW7 2BZ, U.K. }

\address{${}^{b}$Dipartimento di Fisica
``E.R.Caianiello'',
Universit\`a di Salerno, \\ INFN, Gruppo Collegato, Salerno and Unit\`a
INFM, Salerno \\  I-84100
Salerno, Italy}
\maketitle

\begin{abstract}
We comment on the paper ``Remarks on flavor-neutrino propagators and
oscillation formulae''  $[$Phys.~Rev.~D ${\bf 64}$, 013011 (2001)$]$.
We show that the
conclusions presented in that paper do not apply to the exact
field theoretical oscillation formulae
obtained in the BV formalism (for three flavors)
which are free from the dependence on arbitrary mass parameters, account
for CP violation and reduce to the
usual quantum mechanical (Pontecorvo)
 three flavor oscillation formulae in the relativistic
limit.
\end{abstract}

\vspace{0.3cm}

P.A.C.S.: 14.60.Pq

$$  $$

\section{Introduction}

In recent years exact formulae for neutrino
oscillations have
been obtained in the quantum field theory (QFT)
framework\cite{BV95,BHV99,remarks,currents,cinareport}
(hereafter referred to
as the BV formalism
by following Ref. \cite{fujii}). In
the paper ``Remarks on flavor-neutrino propagators and
oscillation formulae''  $[$Phys.~Rev.~D ${\bf 64}$, 013011 (2001)$]$
\cite{fujii2}
it has been remarked that by the use of the retarded propagators different
formulae can be obtained which are free from the dependence on arbitrary
mass parameters and in the case of three flavor oscillations they reduce
to the formulae obtained in the BV formalism only when the mixing matrix is
real. No problem arises in the two flavor case, in the sense that the
retarded propagator derivation gives same oscillation formulae as
obtained in \cite{BHV99}.

In the following we observe that the formulae obtained in Ref. \cite{fujii2}
are not physically acceptable since they do not allow CP violation and
do not reduce to the usual Pontecorvo three flavor formulae in the
relativistic limit. We note
that, on the contrary, our formulae do account for CP
violation, are independent of arbitrary mass parameters and do reduce
to the usual Pontecorvo three flavor formulae in the
relativistic limit.

Apart from the latter property, which by itself is not at all a
negligible requirement to be satisfied, the consistency with CP
violation discriminates with a clear cut between our formulae and
those derived by the use of the retarded propagator. Moreover, it
also points to the necessity of using the flavor Hilbert space, as
indeed emerges in our treatment. In this respect, we observe that
the authors of Ref. \cite{fujii2} actually do not exclude the
possibility of using such a flavor state space. On the contrary, they
analyze the arguments presented in Ref. \cite{GKL92} and conclude
that the assertion there presented against the flavor space is
not appropriate.

The paper is organized as follows. We first introduce the main
lines of our derivation for the three flavor case (see also
\cite{cinareport}). Subsequently, we discuss the results of Ref.
\cite{fujii2} and see that they are physically not acceptable for
the reasons said above.

The derivation of the three flavor oscillation
formulae proceeds along the
same line of the derivation for the two flavor case. We use
standard QFT for the neutrino fields and the familiar
parameterization of the CKM matrix\cite{CKM}. We write down the
generator for the mixing transformations in terms of the Dirac
neutrino fields $\nu_{i}$ with masses $m_{i}$,  $i=1,2,3$,
then we consider the charges for
$\nu_{i}$ and for the flavor neutrinos $\nu_{\sigma}, \sigma = e,
\mu , \tau$. We construct the flavor state space, which is found to
be unitarily
inequivalent to the state space for the mass eigenstate neutrinos,
and we compute the expectation values of the flavor charges in the
flavor states, thereby obtaining the exact oscillation formulae.
We finally discuss the CP violation. Besides the obvious higher
level of computational complexity, the essential, non-trivial
difference with respect to the two flavor case is indeed in the
physically relevant fact that the three flavor oscillation
formulae  must account for CP violation.

We then analyze the derivation of the formulae given in
\cite{fujii2} and discuss their independence of the CP violating
phase and their failure in reducing to the usual Pontecorvo
formulae in the relativistic case.

We observe that the conclusions reached in the present paper can be
extended as well to the case of
boson mixing\cite{lathuile,bosonmix} with number
of flavors larger than two.

We consider the CKM matrix for neutrinos:
\bea\label{fermix} \Psi_f(x) \, =\begin{pmatrix}
  c_{12}c_{13} & s_{12}c_{13} & s_{13}e^{-i\delta} \\
  -s_{12}c_{23}-c_{12}s_{23}s_{13}e^{i\delta} &
  c_{12}c_{23}-s_{12}s_{23}s_{13}e^{i\delta} & s_{23}c_{13} \\
  s_{12}s_{23}-c_{12}c_{23}s_{13}e^{i\delta} &
  -c_{12}s_{23}-s_{12}c_{23}s_{13}e^{i\delta} & c_{23}c_{13}
\end{pmatrix}\,\Psi_m (x) \, , \eea
with $c_{ij}=\cos\theta_{ij},  s_{ij}=\sin\theta_{ij}$ being
$\theta_{ij}$ the mixing angles,
$\Psi_m^T=(\nu_1,\nu_2,\nu_3)$ and
$\Psi_f^T=(\nu_e,\nu_{\mu},\nu_{\tau})$.
The CKM matrix is generated as\cite{BV95}
\bea\
\nu_{\si}^{\alpha}(x)&=&G^{-1}_{\theta}(t)\, \nu_{j}(x)\, G_{\theta}(t)
\eea
where $(\si,j)=(e,1), (\mu,2), (\tau,3)$ and
\bea\
G_{\bf \te}(t)&=&G_{23}(t)G_{13}(t)G_{12}(t),
\\[2mm]
G_{12}(t)&=&\exp\lf[\theta_{12} \int
d^{3}x(\nu_{1}^{\dag}(x)\nu_{2}(x)-\nu_{2}^{\dag}(x)\nu_{1}(x))\ri]
\\
G_{23}(t)&=&exp\lf[\theta_{23}\int
d^{3}x(\nu_{2}^{\dag}(x)\nu_{3}(x)-\nu_{3}^{\dag}(x)\nu_{2}(x)) \ri]
\\
G_{13}(t)&=&exp\lf[\theta_{13}\int
d^{3}x(\nu_{1}^{\dag}(x)\nu_{3}(x)e^{-i\delta}-\nu_{3}^{\dag}(x)
\nu_{1}(x)e^{i\delta}) \ri]\,.
\eea

The free fields  $\nu_i$ (i=1,2,3) can be quantized in the usual
way\cite{IZ} and expressed as (we use $t\equiv x_0$):
\bea\label{2.2} \nu_{i}(x) = \sum_{r} \int d^3 k \lf[u^{r}_{{\bf
k},i} \al^{r}_{{\bf k},i}(t)\:+    v^{r}_{-{\bf k},i}\bt^{r\dag
}_{-{\bf k},i}(t)   \ri] e^{i {\bf k}\cdot{\bf x}} , \eea
with $\al^{r}_{{\bf k},i}(t)=e^{-i\om_{k,i} t}\al^{r}_{{\bf
k},i}(0)$, $\bt^{r}_{{\bf k},i}(t)=e^{-i\om_{k,i} t}\bt^{r}_{{\bf
k},i}(0)$ and  $\om_{k,i}=\sqrt{{\bf k}^2+m_i^2}$. The vacuum for
the mass eigenstates is denoted by $|0\rangle_{m}$:  $\; \;
\al^{r}_{{\bf k},i}|0\rangle_{m}= \bt^{r }_{{\bf
k},i}|0\rangle_{m}=0$.   The anticommutation relations are the
usual ones; the wave function orthonormality and completeness
relations are those of Ref.\cite{BV95}.

The flavor vacuum is  defined as
\bea
|0(t)\rangle_{f}&=&G_{\te}^{-1}(t)\,|0\rangle_{m} \, ,
\eea
and its orthogonality (unitary inequivalence) in the infinite volume
limit to the vacuum $|0\rangle_{m}$ is obtained as shown in
Ref.\cite{BV95}. We observe that the unitary inequivalence of the
flavor vacuum to the mass eigenstate vacuum has been rigorously proved
for the general case of any number of flavors in Ref. \cite{hannabus}.

The flavor fields are expanded as:
\begin{eqnarray}\label{exnuf1}
\nu_{\sigma}(x)     &=& \sum_{r} \int d^3 k
\left[ u^{r}_{{\bf k},j}
\alpha^{r}_{{\bf k},\sigma}(t) +    v^{r}_{-{\bf k},j}
\beta^{r\dag}_{-{\bf k},\sigma}(t) \right]  e^{i {\bf k}\cdot{\bf x}}\,,
\end{eqnarray}
where $\alpha^{r}_{{\bf k},\sigma}(t)\equiv G^{-1}_{\theta}(t)\,
\alpha^{r}_{{\bf k},j}(t)\, G_{\theta}(t)$ and
$\beta^{r\dag}_{-{\bf k},\sigma}(t) \equiv
G^{-1}_{\theta}(t)\,\beta^{r\dag}_{-{\bf k},j} \, G_{\theta}(t)$ with
$(\si,j)=(e,1), (\mu,2), (\tau,3)$.

In order to derive oscillation formulae, we define
the flavor charges $ Q_\si$ ($\sigma = e, \mu, \tau$)
as\cite{currents,cinareport}:
\bea Q_\si(t) & = & \sum_{r} \int d^3 k\lf( \al^{r\dag}_{{\bf
k},\si}(t) \al^{r}_{{\bf k},\si}(t)\, -\, \bt^{r\dag}_{-{\bf
k},\si}(t)\bt^{r}_{-{\bf k},\si}(t)\ri)\,, \eea
which  are connected to the (conserved) Noether charges
$Q_i$ of the free fields
via the mixing generator: $Q_\si(t)
= G^{-1}_{\te}(t)Q_j G_{\te}(t)$, ($(\si,j)=(e,1), (\mu,2), (\tau,3)$).
As usual in QFT, one must perform
subtraction of the vacuum contributions, or, in other words,
use normal ordering with
respect to the vacuum where one operates with charges and currents.

We define the $\rho$-flavor
neutrino state with a given momentum and helicity
as $| \nu_\rho\rangle \equiv \,\alpha^{r \dag}_{{\bf
k},\rho} | 0\rangle_{f}$ and similarly for antiparticles.
In the following for simplicity we use $\al_\rho\equiv
\al^{r}_{{\bf k},\rho}$ and $\bt_\rho\equiv\bt^{r}_{-{\bf k},\rho}$. We then
obtain the oscillation formulae for neutrinos and antineutrinos as
\begin{eqnarray} \label{charge1}
{\cal Q}^\rho_\si(t)&\equiv&
\lan \nu_\rho|Q_\si(t)| \nu_\rho\ran\,
-\,{}_f\lan 0|Q_\si(t)| 0\ran_f = \,
\left|\left\{\alpha_{\sigma}(t), \alpha^{\dag}_{\rho}(0)
\right\}\right|^{2}
\;+ \;\left|\left\{\beta_{\sigma}^{\dag}(t),
\alpha^{\dag}_{\rho}(0) \right\}\right|^{2}\, ,
\\[2mm] \label{charge2}
{\cal Q}^{\bar \rho}_\si(t)&\equiv&
\lan {\bar \nu}_\rho|Q_\si(t)| {\bar \nu}_\rho\ran\,
-\,{}_f\lan 0|Q_\si(t)| 0\ran_f = \,-
\left|\left\{\bt_{\sigma}(t), \bt^{\dag}_{\rho}(0)
\right\}\right|^{2}
\;- \;\left|\left\{\al_{\sigma}^{\dag}(t),
\bt^{\dag}_{\rho}(0) \right\}\right|^{2}\, .
\end{eqnarray}
with $\sum_\si {\cal Q}^\rho_\sigma (t)=
- \sum_\si {\cal Q}^{\bar \rho}_\sigma (t) =1$.

The above formulae coincide with the usual quantum-mechanical
ones in the relativistic limit\cite{cinareport}. Indeed, in this limit, the
(anti-)neutrino state reduces to the usual quantum-mechanical one,
defined on the
vacuum $|0\ran_m$ and
one has\cite{cinareport}  (considering neutrinos for example):
\bea
{\cal Q}^\rho_\si(t)
\longrightarrow
\left|\left\{\alpha_{\sigma}(t), \alpha^{\dag}_{\rho}(0)
\right\}\right|^{2} \, = \,
\lan \nu_\rho|N_\si(t)| \nu_\rho\ran\, =
|\lan \nu_\rho (t)| \nu_\rho(0)\ran|^2\,, \qquad
for ~~|k|\gg \, m_i\; ,\;\; i=1,2,3 \,,
\eea
which is interpretable as a transition probability.

The $CP$ and $T$ violations are calculated as
\cite{cinareport}:
\bea
\De_{\CP}^{\rho \si}(t)
& \equiv& {\cal Q}^\rho_\si(t) +{\cal Q}^{\bar \rho}_\si (t)
\\ [2mm]
\De_{\T}^{\rho \si}(t)
& \equiv& {\cal Q}^\rho_\si(t) -{\cal Q}^{\rho}_\si (- t)
\eea
with $\rho,\si=e,\mu,\tau$ and
$\De_{\CP}^{\rho \si}(t)=\De_{\T}^{\rho \si}(t) \neq 0$ when $\delta
\neq 0$ and $\rho\neq\si$.

In Ref. \cite{fujii} it was noticed that  expanding the
flavor fields in the same basis as the (free) fields with definite
masses (cf. (\ref{exnuf1})) is actually a special choice,
and that a more general possibility exists.
In other words, in the expansion Eq.
(\ref{exnuf1}) one could  use
eigenfunctions with arbitrary masses $\mu_\sigma$, and therefore not
necessarily the same as the masses which appear in the Lagrangian.  On
this basis, the authors of Ref.\cite{fujii} have generalized the BV
formalism by writing the flavor fields as
\begin{eqnarray}\label{exnuf2}
\nu_{\sigma}(x)     &=& \sum_{r} \int d^3 k   \left[
u^{r}_{{\bf k},\sigma} {\widetilde \alpha}^{r}_{{\bf k},\sigma}(t) +
v^{r}_{-{\bf k},\sigma} {\widetilde \beta}^{r\dag}_{-{\bf k},\sigma}(t)
\right]  e^{i {\bf k}\cdot{\bf x}} ,
\end{eqnarray}
where $u_{\sigma}$ and $v_{\sigma}$ are the helicity eigenfunctions with
mass $\mu_\sigma$. We denote by a tilde the generalized flavor operators
introduced in Ref.\cite{fujii} in order to distinguish them from the
ones in Eq.(\ref{exnuf1}).  The expansion Eq.(\ref{exnuf2}) is
more general than the one in Eq.(\ref{exnuf1}) since the latter
corresponds to the particular choice $\mu_e\equiv m_1$, $\mu_\mu \equiv
m_2$. Of course, the flavor fields in Eq.(\ref{exnuf2}) and Eq.(\ref{exnuf1})
are the same fields.
The relation, given in Ref.\cite{fujii}, between the general flavor
operators  and the BV ones is
\begin{eqnarray}\label{FHYBVa}
&&\left(\begin{array}{c}
{\widetilde \alpha}^{r}_{{\bf k},\sigma}(t)\\
{\widetilde \beta}^{r\dag}_{{-\bf k},\sigma}(t)
\end{array}\right)
\;=\; J^{-1}_{\mu}(t)  \left(\begin{array}{c} \alpha^{r}_{{\bf
k},\sigma}(t)\\ \beta^{r\dag}_{{-\bf k},\sigma}(t)
\end{array}\right)J_{\mu}(t) ~~,
\\ [2mm]\label{FHYBVb}
&&J_{\mu}(t)\,=\, \prod_{{\bf k}, r}\, \exp\left\{ i
\mathop{\sum_{(\sigma,j)}} \xi_{\sigma,j}^{\bf k}\left[
\alpha^{r\dag}_{{\bf k},\sigma}(t)\beta^{r\dag}_{{-\bf k},\sigma}(t) +
\beta^{r}_{{-\bf k},\sigma}(t)\alpha^{r}_{{\bf k},\sigma}(t)
\right]\right\}\,.
\end{eqnarray}
with $(\sigma,j)=(e,1) , (\mu,2), (\tau,3)$,
$\xi_{\sigma,j}^{\bf k}\equiv (\chi^{\bf k}_\sigma - \chi^{\bf k}_j)/2$ and
$\cot\chi^{\bf k}_\sigma = |{\bf k}|/\mu_\sigma$,
$\cot\chi_j^{\bf k} = |{\bf k}|/m_j$.
For $\mu_\si\equiv m_j$, one has $J_{\mu}(t)=1$.

As already noticed in Ref.\cite{remarks}, the flavor charge
operators are the Casimir
operators for the Bogoliubov transformation (\ref{FHYBVa}), i.e.
they
are free from arbitrary mass parameters : ${\wti Q}_\si(t) =Q_\si(t) $. This is
obvious also from the fact that they can be expressed in terms of flavor fields
(see Ref.\cite{cinareport}).

Physical quantities should not
carry any dependence
on the $\mu_\si$: in the two--flavor case, it has been shown
\cite{remarks} that the expectation values of the flavor
charges on the neutrino
states are free from the arbitrariness. For three generations, the question is
more subtle due to the presence of the CP violating phase. Indeed, in
Ref.\cite{fujii2} it has been found that the corresponding
generalized quantities depend on arbitrary mass parameters.
However, we find that:
\bea
{\wti {\cal Q}}^{\rho}_\si(t)&=&
\left|\left\{{\wti \alpha}_{\sigma}(t), {\wti \alpha}^{\dag}_{\rho}(0)
\right\}\right|^{2}
\;+ \;\left|\left\{{\wti \beta}_{\sigma}^{\dag}(t),
{\wti \alpha}^{\dag}_{\rho}(0) \right\}\right|^{2}\,=
\,{\cal Q}^{\rho}_\si(t) + F(\mu_\rho,t)
\\ [2mm]
{\wti {\cal Q}}^{\bar \rho}_\si(t)&=&
\,-
\left|\left\{{\wti \bt}_{\sigma}(t), {\wti \bt}^{\dag}_{\rho}(0)
\right\}\right|^{2}
\;- \;\left|\left\{{\wti \al}_{\sigma}^{\dag}(t),
{\wti \bt}^{\dag}_{\rho}(0) \right\}\right|^{2}\, =
{\cal Q}^{\bar \rho}_\si(t) + F(\mu_\rho,t)
\eea
where $F(\mu_\rho, t)$, whose explicit form we do not report here
for sake of shortness, goes to zero  at $t=0$. It
can be shown that $F(\mu_\rho, t)$ vanishes
for $\de=0$ and/or $\mu_\rho=m_j$,
$(\rho,j)=(e,1) , (\mu,2), (\tau,3)$.  This proves that
the invariant (physical) quantities
in the generalized theory are ${\wti {\cal Q}}^{\rho}_\si(t) -
F(\mu_\rho, t)$
and $ {\wti {\cal Q}}^{\bar \rho}_\si(t) - F(\mu_\rho, t)$ which
in fact coincide with the
oscillation formulae of Eqs.(\ref{charge1}) and (\ref{charge2}).
In order to understand the origin of such an invariance,
it would be useful to obtain $F(\mu_\rho, t)$ as
expectation value of some
operator. But this goes beyond the task of the present paper.

We remark that the quantities proposed in Ref.\cite{fujii2}
as probabilities
for flavor oscillations are ruled out by the present analysis. Indeed,
it was there shown that by using the retarded propagators one could
arrive at the
same oscillation formulae of the BV formalism. However,
this coincidence holds only for two flavors since in the three flavor
case, the
quantity (Eq.(3.7) in Ref\cite{fujii2}):
\bea \non
&&P^{(ret)}_{\nu_\si\rar\nu_\rho}(k;t)
\equiv \frac{1}{4} {\rm Tr}[g^{(ret)}_{\rho \si}(k;t)
g^{(ret)\dag}_{\rho \si}(k;t)]
\\
&&= \frac{1}{4} \sum_r\lf[
\left|\left\{\alpha_{\rho}(t), \alpha^{\dag}_{\si}(0)
\right\}\right|^{2}
\;+ \;\left|\left\{\beta_{\rho}^{\dag}(t),
\alpha^{\dag}_{\si}(0) \right\}\right|^{2}
+ \left|\left\{\alpha_{\rho}(t), \beta_{\si}(0)
\right\}\right|^{2}
\;+ \;\left|\left\{\beta_{\rho}^{\dag}(t),
\beta_{\si}(0) \right\}\right|^{2}
\ri]
\eea
is manifestly CP invariant and thus cannot be used as a correct
 QFT generalization
of the QM oscillation formulae. This is, on the other hand, also
confirmed by the fact that the relativistic limit of the above
expression does not give the three flavor Pontecorvo formulae.
On the contrary, we have shown
 that our formulae
do exhibit all the expected features in the presence of a CP violating
phase and Pontecorvo formulae are recovered in the relativistic limit.

Finally, we remark that the formulae Eqs.(\ref{charge1}),(\ref{charge2})
can be also
obtained  by use of the unordered Green's functions \cite{BHV99}
as follows (Eq.(2.37) in Ref.\cite{fujii2}):
\bea
\left|\left\{\alpha_{\rho}(t), \alpha^{\dag}_{\si}(0)
\right\}\right|^{2}
\;+ \;\left|\left\{\beta_{\rho}^{\dag}(t),
\alpha^{\dag}_{\si}(0) \right\}\right|^{2}\, = \,
\frac{1}{2} {\rm Tr}[{\cal G}^{>}_{\rho\si}(k;t)
 {\cal G}^{>\dag}_{\rho\si}(k;t)]
\eea
In Ref.\cite{fujii2}, it is observed that these quantities are
generally dependent on
the arbitrary mass parameters (in the case one calculate
the propagators within
the generalized framework) and also that they can be interpreted as (oscillation)
probabilities since they satisfy the required boundary conditions.
We have already shown how to renormalize the above quantities with respect to the
arbitrary mass parameters. As for the probabilistic interpretation, we need to
stress that flavor states are essentially multiparticle ones and thus one
cannot really talk of probabilities for the evolution of such states. Rather,
the correct interpretation of our exact oscillation formulae is the one
of Eqs.(\ref{charge1}),(\ref{charge2}), i.e. as expectation values of the
flavor charges on states defined on the flavor Hilbert space.
This is evident in the bosonic case, where the corresponding
quantities can assume values larger than one as well as
negative values  \cite{bosonmix,Ji2}.

\section*{Acknowledgements}

This work has been supported by MURST and INFN. Partial support from
 INFM, EPSRC and ESF is also acknowledged.


\end{document}